\documentclass[prc,onecolumn,tightenlines,12pt]{revtex4}
\usepackage{graphicx}
\usepackage{dcolumn}
\usepackage{amsmath}
\begin{document}

\title{Disordered ensembles of random matrices}

\author{O. Bohigas$^{1}$, J. X. de Carvalho$^{2,3}$ and M. P. Pato$^{1,2}$}

\affiliation{
$^{1}$CNRS, Universit\'e Paris-Sud, UMR8626, \\
LPTMS, Orsay Cedex, F-91405, France \\
$^{2}$Inst\'{\i}tuto de F\'{\i}sica, Universidade de S\~{a}o Paulo\\
Caixa Postal 66318, 05315-970 S\~{a}o Paulo, S.P., Brazil\\
$^{3}$Max-Planck-Institut f\"ur Physik komplexer Systeme\\
N\"othnitzer Stra$\beta$e 38, D-01187 Dresden, Germany}

\begin{abstract}
It is shown that the families of generalized matrix ensembles recently
considered which give rise to an orthogonal invariant stable L\'{e}vy ensemble 
can be generated by the simple procedure of dividing Gaussian
matrices by a random variable. The nonergodicity of this kind of
disordered ensembles is investigated. It is shown that the same procedure
applied to random graphs gives rise to a family that interpolates between 
the  Erd\"{o}s-Renyi and the scale free models.
\end{abstract}
\maketitle

The classes of random matrix ensembles introduced by Wigner in the 50s 
have found a great sucess partly after being connected with quantum
manifestations of chaos in physical systems\cite{Boh82}. In turn this
success generated a great activity and extensions 
and generalizations of those ensembles have occurred. In obtaining the
Gaussian ensembles, Wigner adapted the Wishart ensembles 
well known to statistitians. Some of the extensions of the Gaussian 
ensembles can also be considered as applications of known processes in 
statistics. For instance, models to describe symmetry breaking have been 
constructed by adding two random matrices, one block 
diagonal and the other its complement\cite{Guhr}. Here we 
consider a random process in which a new random quantity is generated 
by taking not the sum but the ratio or the product of two other 
independent ones.

In a previous paper\cite{Bertuola}, an alternative 
to Shannon information entropy, namely Tsallis-Renyi information\cite{Tsallis}
was used to introduce a new family of generalized matrix
ensembles (see also \cite{Raul}). One of the main features of this ensemble 
is the power-law characteristic of its statistical properties. In
particular, it was shown that individual matrix elements behave 
like the elements of the so-called L\'{e}vy matrices\cite{Cizeau} 
(after the publication of Ref. \cite{Bertuola}, Klauder and Muttalib 
obtained an even more general family\cite{Klauder} on similar lines). 

One of the purposes of this note is to show that all these families 
can be obtained, in fact, by the following simple procedure. Let
$ H_G (\alpha) $ be a random  matrix of dimension $N$ and 
variance $1/2\alpha ^2$ and let its probability distribution be

\begin{equation}
P_{G} (H ;\alpha )=\left(\frac{\beta\alpha}{\pi}\right)^{f/2}
\exp\left(-\alpha\beta \mbox{tr} H^{2}\right) , \label{12}
\end{equation}
The matrices of the Gaussian ensemble are specified by $\alpha.$
In (\ref{12}), $f$ is the number of independent matrix elements
$f=N+\beta N(N-1)/2$ and $\beta$ is the Dyson index $\beta=1,2,4$ 
for GOE, GUE and GSE (here and in what follows the subindex $G$ 
indicates Gaussian). The distribution is normalized with respect 
to the measure 
$dH=\prod_{1}^{N}dH_{ii}\prod_{j>i}\prod_{k=1}^{\beta}\sqrt{2}dH^{k}_{ij}.$  

Take now a positive random variable $\xi$  with a normalized 
density probability distribution $w (\xi )$ with average $ \bar{\xi}$
and variance $\sigma_{\xi}^2 $ and introduce a new matrix ensemble by
the following relation (product of random variables has been
considered in the context of covariace matrices\cite{Biroli})  

\begin{equation}
H(\alpha, \xi )= \frac{ H_G (\alpha)}{\sqrt{\xi/\bar \xi }} . \label{1}
\end{equation} 
In this way, an external source  of randomness is superimposed to the 
fluctuations of the Gaussian matrix $H_G (\alpha).$ A random process 
in which there is a competition between two types of random 
variables is typical of disordered systems or, in the case of Ising models,
spin glasses\cite{Mezard}. As the two types of randomness are independent, one
can be kept frozen, quenched in technical terms, while the fluctuations 
of the other continue to operate. Here the disorder is represented
by $\xi$ which is the quenched variable in opposition to the
randomness of the Gaussian matrices. We may refer to (\ref{1}) as
a disordered ensemble. 

From (\ref{1}), we deduce that the joint distribution of a set 
of $n\le f$ matrix elements is given by 

\begin{equation}
p(h_1 , h_2 ,..., h_n ;\alpha )=
(\frac{\beta\alpha}{\pi\bar{\xi}})^{n/2}
\int d\xi w (\xi ) \xi^{n/2}\exp\left(-\frac{\beta\alpha\xi}
{\bar{\xi}}\sum_{i=1}^{n} h_i^2  \right) \label{6}  
\end{equation}
where $h_i =H_{ij}$ for the diagonal and $h_i =\sqrt{2}H_{ij}$ 
for the off-diagonal elements. Eq. (\ref{6}) shows that matrix 
elements are correlated. As a particular case, for 
$n=f,$ (\ref{6}) leads to the ensemble distribution 

\begin{equation}
P (H ;\alpha )=
\int d\xi w ( \xi)
\left(\frac{\beta\alpha\xi}{\pi\bar{\xi}}\right)^{f/2}
\exp\left(-\frac{\beta\alpha\xi}{\bar{\xi}}\mbox{tr} H^2  \right) \label{9}  
\end{equation}
where the term after $w(\xi)$ is just (\ref{12}) with $\alpha$
replaced by $\alpha \xi/\bar{\xi}. $ Expressions like (\ref{9}) are
being considered as instances of superstatistics\cite{Abul}.

The relation (\ref{1}) makes straightforward to do numerical
simulations in terms of Gaussian matrices. However, it may also
be useful to directly generate matrices of the 
ensemble (taking into account the corrrelations among their
elements). This can be done through the identity

\begin{equation}
p(h_{1},...,h_{f})=p(h_{1}) \prod_{n=2} ^{f} 
\frac{p(h_{1},...h_{n})}{p(h_{1},...h_{n-1})}, \label{514}
\end{equation}
where each fraction gives the conditional probability for the $n$th
element once the $n-1$ previous ones are given. This equation 
provides a way to sequentially generate all the matrix elements. 
At each step, a new element, say the $n$th, is sorted using 
Eq. (\ref{6}) that implies 
 
\begin{equation}
h_n= \frac{ h_G (\alpha)}{\sqrt{\xi_n /\bar \xi }},  \label{11}
\end{equation} 
where $h_G $ is a Gaussian variable and $\xi_n$ is another random
variable sorted from the distribution

\begin{equation}
w_n (\xi)= w (\xi ) \xi^{(n-1)/2}\exp\left(-\frac{\beta\alpha\xi}
{\bar{\xi}}\sum_{i=1}^{n-1} h_i^2  \right)/
\int d\xi w (\xi ) \xi^{(n-1)/2}\exp\left(-\frac{\beta\alpha\xi}
{\bar{\xi}}\sum_{i=1}^{n-1} h_i^2  \right), \label{5}  
\end{equation}
which is univariate since all the previous $n-1$ elements have 
already been determined. 

By generating matrices fixing, in the process, a set of values  
$\xi_1 , \xi_2 ,...,\xi_f $ we are, in the language of the 
disordered systems, quenching the disorder. The differences among
matrices generated with different sets of $\xi$ depend on the width 
of the distribution $w(\xi)$ and one can expect that for wide $w(\xi)$ 
the large spread among the matrices will give rise to a nonergodic
behavior.     

Turning now to eigenvalues and eigenvectors, we observe that
we have an ensemble invariant
under unitary transformation in which, as it occurs with the 
Gaussian ensembles, the joint 
distribution of  eigenvalues and eigenvector factorizes.
The eigenvectors behave as those of the Gaussian ensembles
and we can integrate them out to obtain for the eigenvalues 
the joint distribution

\begin{equation}
P\left( E_{1},...E_{N};\alpha \right ) =
\int d\xi w \left( \xi \right)(\alpha\xi/\bar{\xi}) ^{\frac{N}{2}}
P_{G}\left( x_{1},...x_{N};\frac{\beta}{2}\right) , 
\label{15}
\end{equation}
where $x_i =\sqrt{\alpha\xi/\bar{\xi} }E_i$ and

\begin{equation}
P_{G} (x_{1},...x_{N}; \frac{\beta}{2} )=
K_{N}^{-1}\exp\left(-\frac{\beta}{2}\sum_{k=1}^{N} x_{k}^{2}\right)
\prod_{j>i} \left| x_{j}-x_{i}\right|^{\beta} ,
\label{26}
\end{equation}
with $K_{N}$ being a normalization constant.

From (\ref{15}), 
measures of the generalized family can be calculated by 
weighting the corresponding measures of the Gaussian ensembles 
with the $w(\xi )$ distribution. Integrating for instance 
(\ref{15}) over all eigenvalues but one and multiplying 
by $N,$ the eigenvalue density is expressed in terms of 
the Wigner's semi-circle law\cite{Meht} as

\begin{equation}
\rho \left( E;\alpha \right) =\frac{\sqrt{2\alpha}} {\pi} 
\int d\xi w( \xi) (\xi/\bar{\xi})^{\frac{1}{2}}\sqrt{2N-2\alpha 
\xi E^{2}/\bar{\xi} }. \label{126}
\end{equation}
where the condition $\alpha \xi E^{2}<N$ on
$\xi$ has to be satisfied.

As previously stated, the introduction of the disorder represented by the
variable $\xi,$ breaks in principle the ergodicity of the Gaussian
ensembles. 
Let $N(L)= \int_{ E -L/2 }^{ E +L/2} dE^{\prime} \rho(E^{\prime})$ 
be the average number of eigenvalues in the interval
$[ E -L/2,  E + L/2]$ for an ensemble with eigenvalue density $\rho(E).$
The variance $ \Sigma^2 (L) $ of the number of eigenvalues in 
that interval can be expressed in terms of the two-point 
correlation function $R(E_1 ,E_2)$ by 

\begin{equation}
\Sigma^2 (L) =  \int_{ E -L/2 }^{ E +L/2} dE_1 
\int_{ E -L/2}^{ E +L/2} 
dE_2 R(E_1 ,E_2) +N(L) - N^2 (L). 
\end{equation}  
Ergodicity implies\cite{Pandey} the vanishing of

\begin{equation}
\mbox{Var} \rho = [\rho( E)]^2 \Sigma^2 (L) /L^2.  \label{16}
\end{equation}  
when $L\rightarrow \infty.$ For the disordered 
ensemble we have 

\begin{equation}
\Sigma^2 \left( L \right) =\int d\xi w
\left(\xi \right) \left[ 
\Sigma^2_G (L) - N_G (L)+ N_G ^2 (L) \right]+ 
N(L) - N^2 (L).\label{37}
\end{equation}
with $N_G (L)$ calculated with the Gaussian density. In
(\ref{37}), nonergodicity will result if the quadratic terms do 
not cancel. Indeed, in this case, a parabolic contribution
for  large $L$ survives and the variance of the density fluctuations
given by Eq. (\ref{16}) does not asymptotically vanish. 

Consider now a particular choice of the distribution
$w(\xi).$ Note that the factor multiplying 
the Gaussian matrices in Eq. (\ref{1}) acts on the variance of the 
Gaussian ensembles. In order to investigate ensembles showing 
heavy-tailed densities it is convenient to choose $w(\xi)$ to be 
the gamma distribution 

\begin{equation}
w(\xi)=
\exp(-\xi) \xi^{\bar{\xi} -1} /\Gamma(\bar{\xi})  \label{18}
\end{equation} 
that becomes a $\chi^2$ distribution for integer $2\bar{\xi}$. 
From (\ref{18}) $\sigma_{\xi}=\sqrt{\bar{\xi}},$ showing that
$\bar{\xi}$ controls the 
behavior of the distribution $w(\xi).$ It becomes more localized 
when  $\bar{\xi}$ increases and we should then expect 
to recover the Gaussian ensembles. However, for smaller 
values of $\bar{\xi},$ departures from the Gaussian case will be observed.
Indeed, by substituting (\ref{18}) in (\ref{9}) we find

\begin{equation}
P(H;\alpha ,\bar{\xi} )=\left(\frac{\beta\alpha}{\pi\bar{\xi}}
\right)^{\frac{f}{2}}\frac
{\Gamma \left( \frac{1}{q-1}\right)}{\Gamma \left( \bar{\xi} \right) }
\left(1+\frac{\beta\alpha}{\bar{\xi}}
\mbox{tr} H^{2}\right) ^{\frac{1}{1-q}}  \label{22}
\end{equation}
for the ensemble density distribution, where

\begin{equation}
 \frac{1}{q-1}=\bar{\xi}+\frac{f}{2} , \text{ with   } q>1.
\end{equation}
Eq. (\ref{22}) is just Eq. (4) of \cite{Bertuola}. 
In \cite{Bertuola} it was derived using  
a generalized maximum entropy principle\cite{Tsallis} with $q$ being 
identified with the Tsallis entropic parameter. 

Substituting (\ref{18}) in (\ref{6}) for $n=1$\cite{Burda}

\begin{equation}
p(h;\alpha,\bar{\xi})=\left(\frac{\beta\alpha}{\pi\bar{\xi}}
\right)^{\frac{1}{2}}\frac{\Gamma \left(\bar{\xi} +1/2 \right) }
{\Gamma \left( \bar{\xi}\right)}
\left(1+\frac{\beta\alpha}{\bar{\xi}}
h^{2}\right) ^{-\bar{\xi}-1/2}  \label{280}
\end{equation}
for the density  distribution of a given matrix element.  
Since for large $\left|h\right|,$ 
$p_{\beta}(h;\alpha,\bar{\xi}) \sim 1/\left|h\right|^{2\bar{\xi}+1},$ 
(\ref{280}) exhibits the power-law character of the distribution.
It is important to remark that, apart from the lack of independence, 
the marginal distribution of the matrix 
elements have the same kind of distribution, namely one with an 
asymptotic power-law behavior, as the i.i.d. ones of the ensemble 
of L\'{e}vy matrices\cite{Cizeau}

In Fig. 1 the eigenvalue density for three realizations of the 
ensemble generated using the above random process with  $\bar{\xi}=1/2$ 
is histogrammed and compared with the semi-circle law. We recall 
that for $\bar{\xi} =1/2$ the matrix elements are Cauchy,
$\frac{1}{\pi}\frac{1}{1+x^2},$  distributed(see Eq. (\ref{280})). 
It is seen that the individual matrices of large sizes are 
Gaussian ensemble matrices as they should. As a comparison, 
in Fig. 2, it is shown the eigenvalue
density of just one L\'{e}vy matrix of large size whose matrix 
elements also follow the
Cauchy distribution. We can see that although individual matrix
elements of the two ensembles are identically distributed, their 
eigenvalue density behaves in a completely different way. While 
individual L\'{e}vy matrices of large sizes do not depart from the
ensemble average, matrices generated according 
to (\ref{11}) show large fluctuations. 

Of course, the result shown in Fig. 1 indicates strong nonergodicity. 
This is confirmed by the ensemble number variances shown in Fig. 3.
The parabolic behavior seems to persist even for large values of 
the parameter $\bar{\xi}, $ showing that the ensemble is nonergodic. 
Consequently, averages performed running along one
spectrum do not coincide with averages over the
ensemble of matrices. 

Other systems in which nonergodicity may play an important role are
networks and their associated graphs. We now show how the present approach 
can be applied in random graph theory\cite{Albert}. 
A graph is an array of points (nodes) connected by edges. It is  
completely defined by its adjacency matrix $A$ whose elements $A_{ij}$ 
have value $1$($0$) if the pair $(ij)$ of nodes is connected (disconnected).
The diagonal elements are taken equal to zero, i.e. $A_{ii}=0.$ 
Adjacency matrices of graphs in which the connections are randomly set, 
are real symmetric random matrices. The classical random graph model 
proposed by  Erd\"{o}s-Renyi (ER) is simply 
defined by giving a fixed 
probability $p$ that a given pair of nodes is connected, independently 
of the others\cite{Renyi}.

We start by showing that the ER model can be considered as the
equivalent in random graph theory to the Wigner model of Gaussian 
matrices. In fact, the joint matrix element distribution of its
adjacency matrix $A$ can be written as

\begin{equation}
P_{ER}(A,\alpha)=\left[1+\exp(-\alpha)\right]^{-f}
\exp\left(-\frac{\alpha}{2} \mbox{tr} A^{2}\right) \label{17}
\end{equation}
where $f=\frac{N(N-1)}{2}$ with
$N,$ the size of matrix, being equal to the number of nodes. Eq. (\ref{17}) 
is just the defining equation  (\ref{12}) of the GOE ($\beta=1$) ensemble
with the constraint that the matrix elements
can only take the values $0$ and $1$ imposed by the measure

\begin{equation}
dH=\prod_{1}^{N}dH_{ii}\delta(H_{ii})\prod_{j>i}
\sqrt{2}dH_{ij}\left[\delta(H_{ij})+\delta(1-H_{ij})\right]. \label{516}
\end{equation} 
From (\ref{17}) it follows that the marginal distribution of a given matrix 
element, say $A_{ij }$, is 

\begin{equation}
P_{ER}(A_{ij},\alpha)=\frac{\exp\left(-\alpha A_{ij}\right)}{1+\exp(-\alpha)} 
=\left\{
\begin{array}{rl}
\frac{\exp(-\alpha)}{1+\exp(-\alpha)} ,& \text{if } A_{ij} = 1 \\
\frac{1}{1+\exp(-\alpha)}, & \text{if } A_{ij} = 0 ,
\end{array}
\right.
\end{equation}
which means that the  probability $p$ that defines the ER model
is connected to the parameter $\alpha$ by the relation 

\begin{equation}
\alpha=\ln(\frac{1}{p}-1).
\end{equation} 
Since the probability $p$ is defined in the interval $[0,1],$ the 
domain of variation of $\alpha$ is $]\infty,-\infty [.$ This suggests
that the statistical properties of the ER model must show a
symmetry with respect to the point $\alpha=0$ (or $p=1/2$).   

It is important to remark that although Eq. (\ref{17}) has the same
structure as Eq. (\ref{12}) there are striking differences between the
two models. Despite the presence 
of the trace in (\ref{17}), the discrete nature of matrix elements 
imposed by the measure, Eq. (\ref{516}), destroys
the rotational invariance and  prevents the factorization
of the joint distribution of eigenvalues and eigenvectors. 
The parameter $\alpha$ is just a scaling parameter in the Gaussian 
case. In contrast, the properties of ER model depend strongly
on the value of the probability $p,$ and here $\alpha$ plays an
essential role. Notice also that, contrarily to
the Gaussian cases, the adjacency matrices form an ensemble with a finite 
number of matrices.    
It is convenient in the study of the graphs, to introduce the scaling 
$p \sim N^{-z}$ ($z>0$). For instance, connectivity properties of the 
graph are characterized by $z.$

An analytical expression of the spectral density for arbitrary values of 
the probability $p$ and matrix size $N$ is an unsolved problem
\cite{Leticia}. However, 
when $p$ is fixed and $N$ is very large, the density can be deduced in
the following way. $A$ is a symmetric non-negative matrix with maximum
principal eigenvalue, $E_1,$ its value is close to 
the nonzero eigenvalue of the constant matrix $<A>$ with elements  
equal to the average of the $A$-elements, i.e. $<A>_{ij}=p.$ 
As the only nonzero eigenvalue of a constant matrix is equal to the product of 
its size by the element, we conclude that $E_1 =pN.$ Because of this linear 
dependence with $N,$ for fixed $p$ the largest eigenvalue grows faster
than the 
others as the matrix size increases. In this case, for very large matrices 
the other eigenvalues have asymptotically the same eigenvalue density of the 
eigenvalues of the matrix $A-<A>.$ This density can be obtained from the 
moments of the trace of the powers of the matrix and one finds that 
it obeys the Wigner semi-circle law\cite{Albert} 

\begin{equation}
\rho_{ER}(E,\alpha)=\left\{
\begin{array}{rl}
\frac{1}{2\pi \sigma^2}\sqrt{4N\sigma^2-E^{2}}, &\mbox{if } 
|E|<\sqrt{4N\sigma^2}\\
0, &\mbox{if } |E|>\sqrt{4N\sigma^2}
\end{array}
\right.
\end{equation}
where $\sigma^2$ is the variance of the matrix elements given by

\begin{equation}
\sigma^2=p(1-p)=\frac{1}{4\cosh^2(\alpha/2)}. 
\end{equation}
The above argument fails if $p\sim 1/N $ ($z\sim 1$) in which case 
deviations from the semi-circle appear\cite{Leticia,Farkas}.

We now introduce a disordered model of random graphs by defining 
an adjacency matrix with a distribution  

\begin{equation}
P(A;\alpha)=\int d \xi 
w(\xi)\frac{\exp\left(-\frac{\alpha\xi}{2} \mbox{tr} A^{2}\right)}
{\left[1+\exp(-\alpha\xi)\right]^{f}}. \label{515}
\end{equation}
Therefore this generalized model is a superposition 
of  Erd\"{o}s-Renyi random graphs with distribution 
$P(A,\alpha \xi)$ weighted with $w(\xi )$ exactly as in (\ref{9}) 
for the disordered Gaussian ensembles. Again 
the width of the distribution of $w(\xi )$ is a controling parameter 
and as remarked before the parameter $\alpha$ also plays an essential 
role. In particular, for $\alpha =0$ the 
ensemble is just the ER with $p=1/2.$

From Eq. (\ref{515}) we can derive the probability distribution for a
set of matrix elements and use Eq. (\ref{514}) to define a random
process entirely equivalent to the one used to generate matrices of
the disordered Gaussian ensemble. As before, a set of probabilities
$p_n$ with $n=1,2,3...,f$ is sequentially generated and, from
them, each new matrix element is obtained taking into account 
those already determined. This means that Eq. (\ref{515}) defines 
a model of a disordered correlated graph in which new attachments
depend on the ones already existing. 

As in the case of the Gaussian ensembles, statistics 
of the averaged graph (our model) are averages over the ER statistics. 
For instance, the eigenvalue density is

\begin{equation}
\rho(E;\alpha)=\frac{2}{\pi}\int^{\xi_m}_{0}d\xi
w(\xi)\cosh(\frac{\alpha\xi}{2})
\sqrt{N-\cosh^2(\frac{\alpha\xi}{2})E^{2}} \label{518}
\end{equation}
where

\begin{equation}
\xi_m=\frac{2}{\alpha}\cosh^{-1} (\frac{\sqrt{N}}{E}).
\end{equation}

We now make for $w(\xi)$ the same choice as before, namely 
Eq. (\ref{18}). 
As before we expect for large values of $\bar{\xi}$ small fluctuations 
around ER, whereas
for small values they will become large and will govern the asymptotics.

In Fig. 4 we display the density of eigenvalues of the adjacency matrices.
When going from $z$ close to 1 to $z$ close to $0,$ the density goes from a
highly picked density with heavy tails towards a Wigner semi-circle, showing
a crossover which is reminiscent from a scale-free to an ER graph.

In summary, we have discussed a new method to introduce matrix ensembles
which preserve unitary invariance presenting distribution with heavy
tails. The price to pay to preserve unitary invariance is i) to abandon the
statistical independence of the matrix elements ii) to abandon the ergodic
property (equivalence of spectral and ensemble averages). 
There are cases, however, in which only ensemble averages make sense.
Consider, for instance, the behavior of individual eigenvalues. 
Recently, extreme eigenvalues have been a matter of great interest due to
the discovery that the distributions they follow, the so-called 
Tracy-Widom\cite{TW} in the case of the Gaussian ensembles, show 
universality and have wide 
applications\cite{TW1}. The same authors have found growing systems 
in which an external source induces the extreme values to have a
behavior in which there is a competition between their distribution and a
Gaussian\cite{Widom}. In a paper in preparation, we show that the
disordered ensemble can be a useful model for this kind of systems.

Let us finally mention that the method discussed here (Eq. (\ref{1}) 
with the choice Eq. (\ref{18}) for the probability density function 
$w(\xi)$) was intended to
rederive and to give new insight on models previously studied. By
making other choices for $w(\xi)$ new models preserving orthogonal
invariance may be introduced (see also \cite{Klauder}).

We thank L. Pastur and W. F. Wreszinski for fruitful discussions.
This work is supported in part by the Brazilian agencies CNPq and FAPESP.

{\bf Figure Captions}

Fig. 1 The eigenvalue density of three matrices of size $N=300$
generated using Eqs. (\ref{11}) and (\ref{18}) with  $\bar{\xi}=1/2$ 
compared with Wigner's semi-circle law. 

Fig. 2 The eigenvalue density of one L\'{e}vy matrix of size $N=600$
whose elements are Cauchy distributed compared to a Cauchy distribution. 

Fig. 3 Full lines: the number variances calculated with Eq. (\ref{37})
for the values $\bar{\xi} =5,10,20,50$ and $200$ as
indicated in the figure; dashed lines: the linear Poisson number
variance and the GOE number variance. 

Fig. 4 The eigenvalue density of the disordered random graph model
calculated with Eqs. (\ref{518}) and (\ref{18}) with $\bar{\xi} =1/2$ 
and for values $ 0.2,$ $ 0.3$ and $0.8$ of the scaling parameter $z.$


\begin{thebibliography}{99}

\bibitem{Boh82} O. Bohigas, M. J. Giannoni, and C. Schmit, Phys. Rev.
Lett. {\bf 52}, 1 (1984); M. Sieber and K. Richter, Physica 
Scripta T {\bf 90}, 128 (2001); S. M\"{u}ller,
S. Heusler, P. Braun, F. Haake and A. Altland, Phys. Rev. E {\bf 72},
046207 (2005); S. Heusler, S. M\"{u}ller, A. Altland, P. Braun, F. Haake, 
Phys. Rev. Lett. {\bf 98}, 044103 (2007).

\bibitem{Guhr}  T. Guhr and, H.A. Weidenm\"{u}ller. Ann. Phys. (NY),
{\bf 199}, 412 (1990); M. S. Hussein and M. P. Pato, Phys. Rev. Lett.
{\bf 70}, 1089 (1993).

\bibitem{Bertuola} A. C. Bertuola, O. Bohigas, and M. P. Pato,
Phys. Rev. E {\bf 70}, 065102(R) (2004).

\bibitem{Tsallis}  C. Tsallis, R. S. Mendes and A. R. Plastino, Physica A 
{\bf 261}, 534 (1998).

\bibitem{Raul}  F. Toscano, R.O. Vallejos, and C. Tsallis,
Phys. Rev. E. {\bf 69}, 066131 (2004);  A. Y. Abul-Magd, 
Phys. Rev. E. {\bf 71}, 066207 (2005).

\bibitem{Cizeau}  P. Cizeau and J. P. Bouchaud, Phys. Rev. E {\bf 50}, 1810
(1994); Z. Burda, R. A. Janik, J. Jurkiewicz, M. A. Nowak, G. Papp,
and I. Zahed, Phys. Rev. E {\bf 65}, 021106 (2002); N. S. Witte and 
P. J. Forrester, Nonlinearity {\bf 13}, 1965 (2000).

\bibitem{Klauder}  K.A. Muttalib and J.R. Klauder,
Phys. Rev. E. {\bf 71}, 055101(R) (2005).

\bibitem{Biroli} G. Biroli, J. P. Bouchaud and M. Potters, 
arXiv:0710.0802v1 [cond-mat.stat-mech]. 

\bibitem{Mezard} M. M\'{e}zard, G. Parisi and M. Virasoro,  
{\it Spin glasses theory and beyond} (World Scientific, Singapore, 1987). 


\bibitem{Abul} C. Beck and E. G. D. Cohen, Physica A {\bf 322}, 267
  (2003); A. Y. Abul-Magd,  Phys. Rev. E {\bf 72}, 066114 (2005).

\bibitem{Meht}  M. L. Mehta, {\it Random Matrices} (Elsevier Academic Press, 
3nd Ed., 2004).

\bibitem{Pandey}  A. Pandey, Ann. Phys. {\bf 119}, 119 (1979).

\bibitem{Burda} Z. Burda, A. T. G\"{o}rlich,and B. Waclaw, 
Phys. Rev. E {\bf 74}, 041129 (2006).

\bibitem{Albert} R. Albert and A.-L. Barab\'{a}si,
  Rev. Mod. Phys. {\bf 74}, 47 (2002).

\bibitem{Renyi} P. Erd\"{o}s and A. Renyi, Publ. Math. Debrecen 
{\bf 6}, 290 (1959).

\bibitem{Leticia} G. Semerjian and L. F. Cugliandolo, J. Phys. A:Math.
  Gen. {\bf 35}, 4837 (2002).

\bibitem{Farkas} I. J. Farkas, I. Der\'{e}nyi, A.-L. Bar\'{a}basi, and 
T. Vicsek, Phys. Rev. E {\bf 64}, 026704 (2001).

\bibitem{TW} C. A. Tracy and H. Widom, Commun. Math. Phys. {\bf 159},
 151 (1994) and  {\bf 177}, 727 (1996).

\bibitem{TW1} C. A. Tracy and H. Widom, Proceedings of the ICM, 
Beijing 2002, vol. 1, 587--596.

\bibitem{Widom} J. Gravner, C. A. Tracy and H. Widom, Ann. of Prob.
{\bf 30}, 1340 (2002); Commun. Math. Phys. {\bf 229}, 433 (2002); 
K. Johansson, Prob. Theo. and Rel. Fields {\bf 138}, 75 (2007). 

\end{thebibliography}
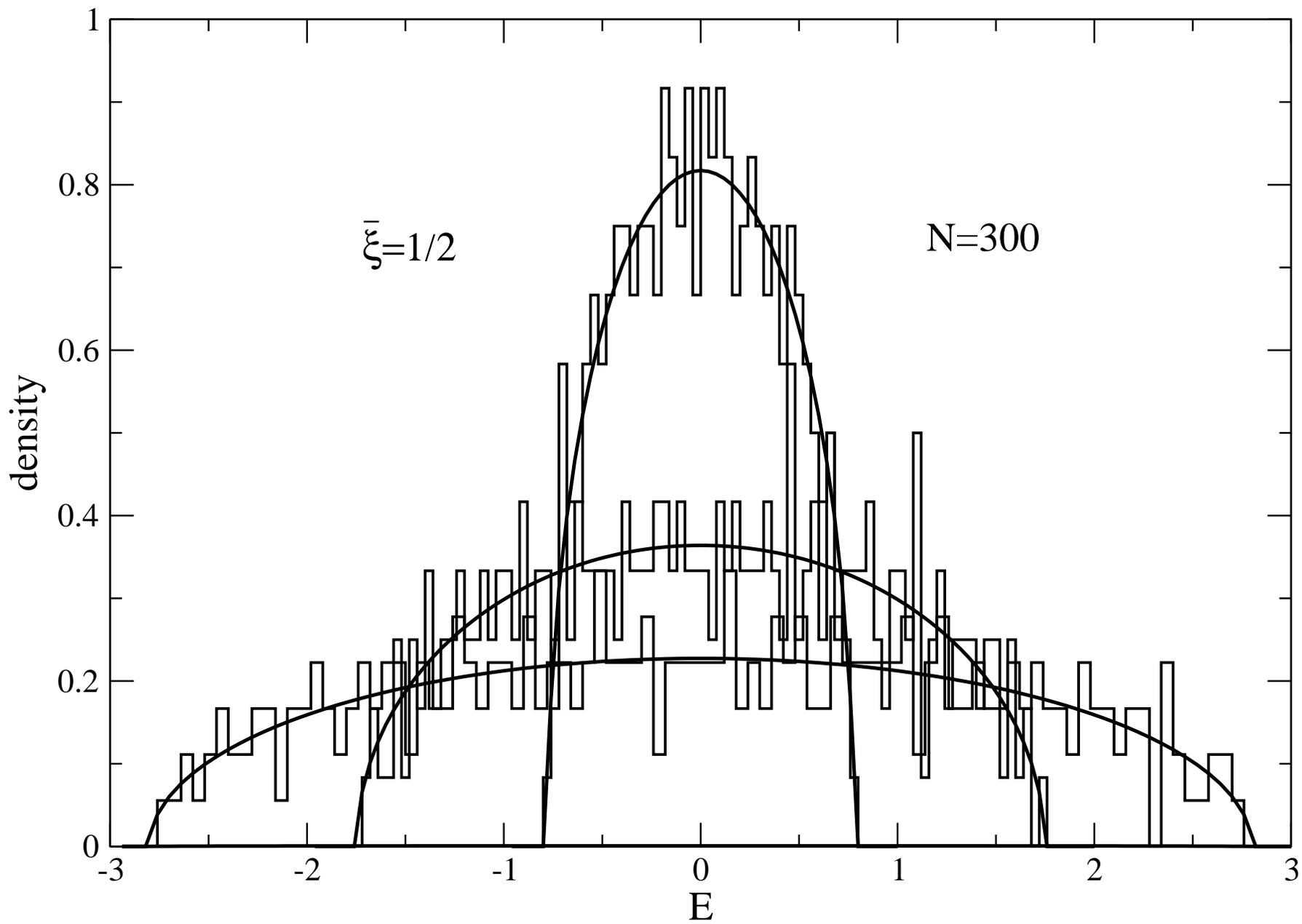
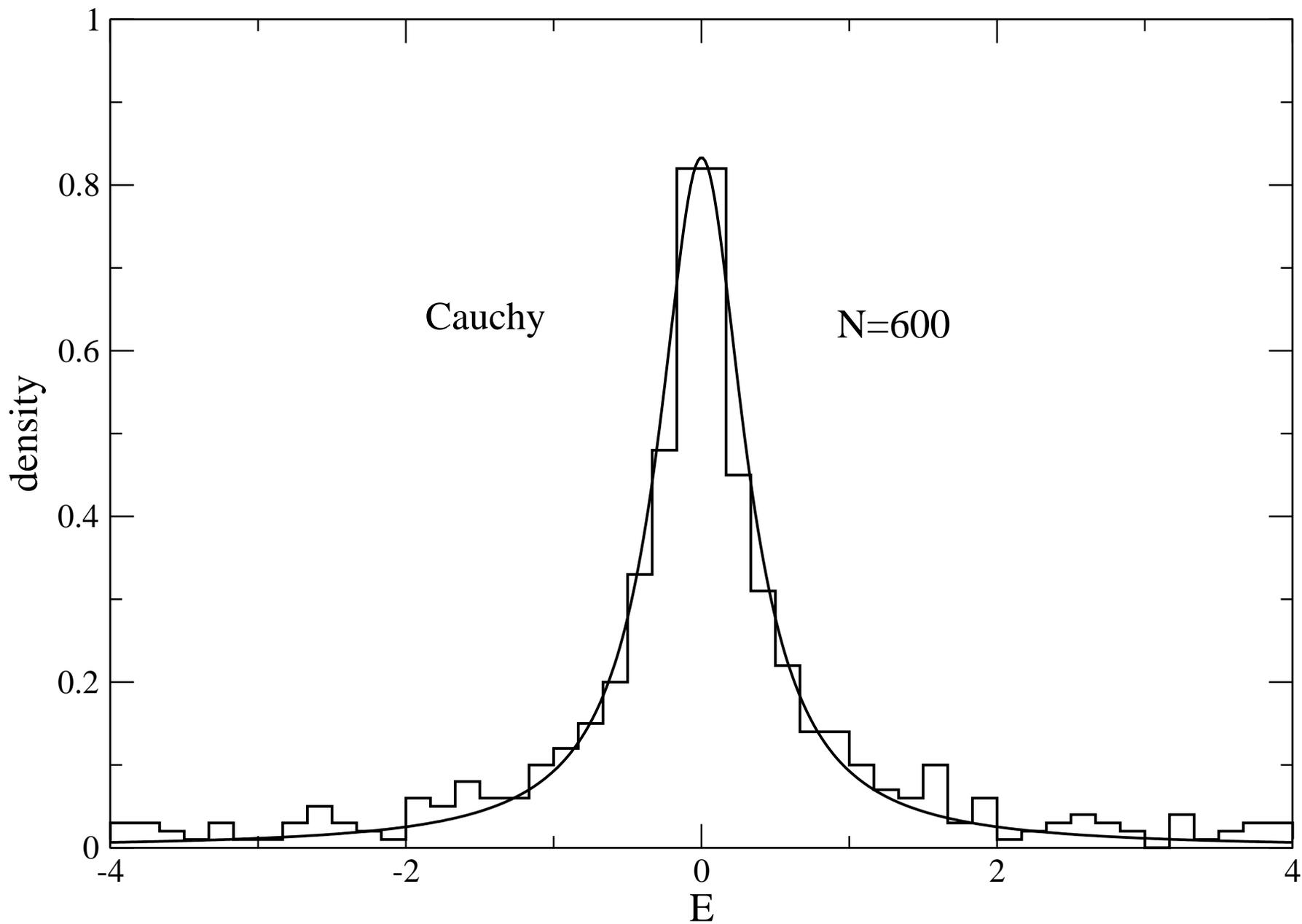
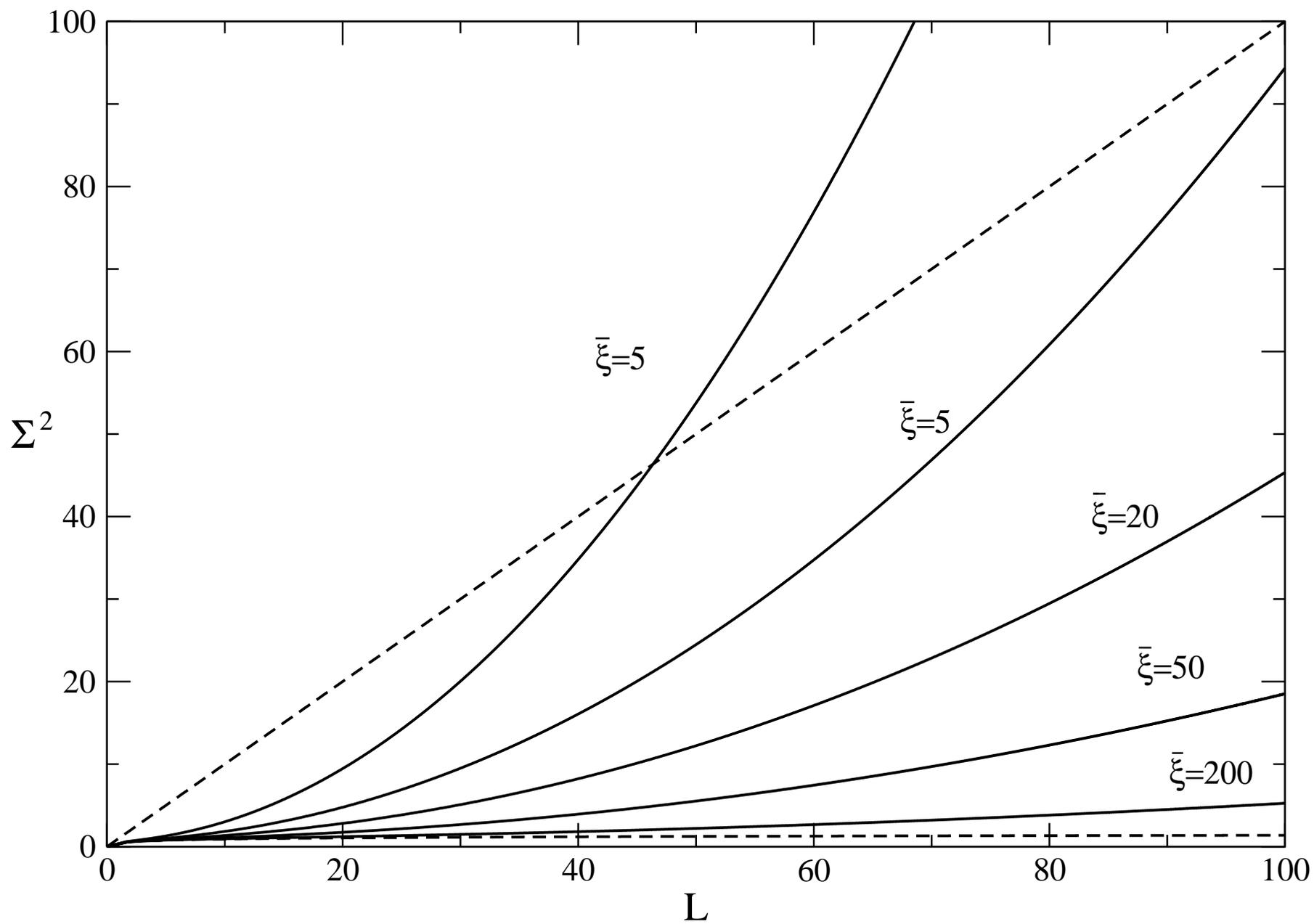
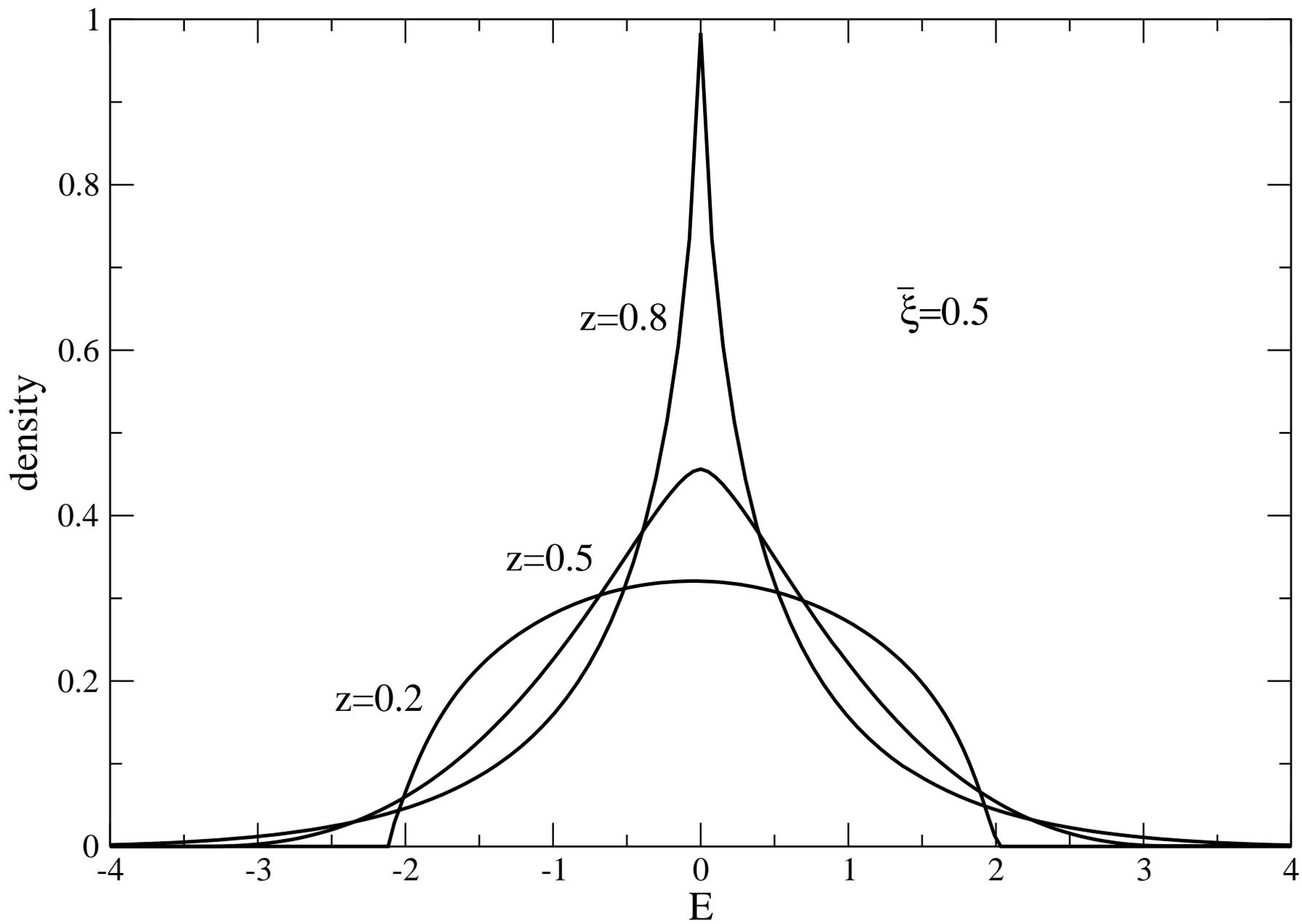
\end{document}